# Sensor Scheduling for Optimal Observability Using Estimation Entropy

Mohammad Rezaeian
Department of Electrical and Electronic Engineering,
University of Melbourne, Victoria, 3010, Australia
Email: rezaeian@unimelb.edu.au.

*Abstract*— We consider sensor scheduling as the optimal observability problem for partially observable Markov decision processes (POMDP). This model fits to the cases where a Markov process is observed by a single sensor which needs to be dynamically adjusted or by a set of sensors which are selected one at a time in a way that maximizes the information acquisition from the process. Similar to conventional POMDP problems, in this model the control action is based on all past measurements; however here this action is not for the control of state process, which is autonomous, but it is for influencing the measurement of that process. This POMDP is a controlled version of the hidden Markov process, and we show that its optimal observability problem can be formulated as an average cost Markov decision process (MDP) scheduling problem. In this problem, a policy is a rule for selecting sensors or adjusting the measuring device based on the measurement history. Given a policy, we can evaluate the estimation entropy for the joint state-measurement processes which inversely measures the observability of state process for that policy. Considering estimation entropy as the cost of a policy, we show that the problem of finding optimal policy is equivalent to an average cost MDP scheduling problem where the cost function is the entropy function over the belief space. This allows the application of the policy iteration algorithm for finding the policy achieving minimum estimation entropy, thus optimum observability.

## I. INTRODUCTION

Sensor scheduling aims to achieve optimum visibility of the sensing process. This problem arises in situations where a number of sensors are set to measure a process, and different sensors provide different visibility depending on the states of the process. However, the sensor management strategies, band limited communications, or the network itself only allows one sensor reading at a time. An example of such a network is a set of sensors deployed densely at a region to track moving targets, and only one can communicate at a time. Depending on the target position, sensors will have different visibility of the target, each providing a vague estimate of the target position. Another example is the waveform selection for a radar system, where various waveforms have different effects for target observation ([1],[2]). In these systems or networks a sensor selection policy needs to be implemented to ensure maximum information flow from the process to the observer. One the other hand, if the parameters of a single measuring (sensing) device can be adjusted where different values of the parameters provide better observability at different states of a process, then finding the optimal on-line adjustment policy can also be considered as a sensor scheduling problem.

Networks of these kinds when the process is a Markov process and measurements are memoryless can be modelled as POMDP. However in contrast to usual applications of POMDPs as a control process problem, here we deal with a different kind of problem that we call it *observability problem*. In such a problem the Markov process is autonomous and the action doesn't influences its evolution. Instead our action influences the visibility of the Markov process which could also depend on its state. Similar to the controllability problem, here also a policy is a rule for choosing the action based on the *belief* on the Markov process. This belief is built up by the measurement history as a variable over an infinite state space. The problem of finding the optimal policy for both of these problems can turn to a Markov decision scheduling problem where the Markov process is the state of belief on its infinite space. In contrast to the controllability problem that the cost function over the belief space is a linear function [3] (irrespective of the cost associations to the states of Markov process), in the observability problem the cost function cannot be a linear function. In the latter problem the aim is to control the belief state to move only between almost sure (low entropy) regions of belief space. In this paper we formulate the observability problem as an average entropy (as cost) MDP scheduling problem.

An average cost MDP scheduling problem finds an optimal policy that minimizes the expected time average cost of the controlled Markov process [4]. When the state space of MDP is finite, value iteration algorithm and policy iteration algorithm (PIA) are simple methods to obtain the optimum scheduling policy. However for general state spaces these algorithms are not applicable or easy to implement. For the linear cost function, the value iteration algorithm has been implemented as an iterative application of linear programming, in particular using incremental pruning [5]. On the other hand, for the average cost MDP scheduling problem with non-linear cost function over general spaces the policy iteration algorithm has been considered as iteratively solving a version of Poisson equation, and the convergence under some conditions has been verified for this algorithm [6].

In this paper the basic relation between POMDP and its corresponding MDP on belief state is captured by introducing H-processes. In the observability problem, for each policy of the controlled H-process the estimation entropy is defined as the limit of conditional entropy of the hidden component given all the past observable component [7]. The effectiveness of a policy can be judged by this entropy measure. We show that under ergodicity condition this limiting entropy is the average cost of the MDP under that policy when the cost function is the entropy function over the belief space. This results in the formulation of the observability problem as an average cost MDP scheduling problem over belief space for which we can apply the results of [6] to implement the policy iteration algorithm for solving optimal policy.

In the next section we discuss advanced results on Markov processes and their controlled version, including ergodicity conditions and Poisson equation. The policy iteration algorithm is also described in this section. In Section III we describe the observability problem using H-processes. The estimation entropy as the cost function for the observability problem is then analyzed in Section IV. In the last section the exact MDP scheduling problem for sensor scheduling adapted to the application of policy iteration algorithm is formalized.

In this paper the domain of a random variable $X$ is denoted by $\mathbb{X}$ if it is a general space, or by $\mathcal{X}$ if it is a finite set. A discrete time stochastic process is denoted by $\mathbf{X} = \{X_n : n \in Z\}$. For a process $\mathbf{X}$, a sequence of $X_0, X_1, ... X_n$ is denoted by $X_0^n$, whereas $X^n$ refers to $X_{-\infty}^n$. The probability $Pr(X = x)$ is shown by $p(x)$ (similarly for conditional probabilities), whereas $\underline{p}(X)$ represents a row vector as the distribution of $X$, ie: the $k$-th element of the vector $\underline{p}(X)$ is $Pr(X = k)$. For a random variable $X$ defined on a set $\mathcal{X}$, we denote by $\nabla_{\mathcal{X}}$ the probability simplex in $\mathbb{R}^{|\mathcal{X}|}$. A specific elements of a vector or matrix is referred to by its index in square brackets. The entropy of a random variable $X$ is denoted by $H(X)$ whereas $h : \nabla_{\mathcal{X}} \to R^+$ represents the entropy function over $\nabla_{\mathcal{X}}$, i.e: $h(\underline{p}(X)) = H(X)$ for all possible random variables $X$ on $\mathcal{X}$.

## II. MARKOV DECISION PROCESSES AND POLICY ITERATION ALGORITHM

A time homogeneous Markov chain $\mathbf{X} = \{X_n : n \in Z\}$ on the general space $\mathbb{X}$ is defined by a conditional probability $P(x, B)$, $x \in \mathbb{X}, B \in \mathcal{B}(\mathbb{X})$. For a given Markov chain and for a cost function $c : \mathbb{X} \to \mathbb{R}_+$, the following functions are defined as discounted cost and average costs, respectively,

$$\begin{aligned} V(x) &= \sum_{t=0}^{\infty} \alpha^t \mathbb{E}[c(X_t)|X_0 = x], \\ J(x) &= \limsup_{n \to \infty} \sum_{t=0}^{n-1} \frac{1}{n} \mathbb{E}[c(X_t)|X_0 = x], \end{aligned} \quad (1)$$

where $X_t$ evolves based on $P$ and $0 < \alpha < 1$ is the discount factor.

A Markov decision process (MDP) is defined by a set of conditional probability $P_a(x, B)$, $x \in \mathbb{X}, B \in \mathcal{B}(\mathbb{X}), a \in \mathcal{A}$, where $\mathcal{A}$ is the control set. A Markov decision process with a control policy $w : \mathbb{X} \to \mathcal{A}$ is a Markov process with $P_w(x, B) = P_{w(x)}(x, B)$. For a given cost function $c : \mathbb{X} \to \mathbb{R}_+$[1], the discounted cost $V(w, x)$ and average costs $J(w, x)$ are also functions of the policy $w$ and are defined by Equation (1) where $X_t$ evolves based on $P_w$. A

[1]In more general settings the cost function is $c : \mathbb{X} \times \mathcal{A} \to \mathbb{R}$. The results are easily extended to this case if $c = c_1 + c_2, c_1 : \mathbb{X} \to \mathbb{R}, c_2 : \mathcal{A} \to \mathbb{R}$.

MDP scheduling problem is to find an optimal policy $w^*$ that minimizes one of these cost criterion, (for all $x \in \mathbb{X}$), [4],

$$\begin{aligned} w^* &= \arg\min V(w, x) \quad \text{Discounted cost problem} \\ w^* &= \arg\min J(w, x) \quad \text{Average cost problem} \end{aligned} \quad (2)$$

Let $B(\mathbb{X})$ be the set of real-valued bounded measurable functions on $\mathbb{X}$, and $\mathcal{M}(\mathbb{X})$ be the set of probability measures on $\mathbb{X}$. For a conditional probability $P$, we define two operations on $B(\mathbb{X})$ and $\mathcal{M}(\mathbb{X})$ as follows [4],

$$Pc(x) \triangleq \int_{\mathbb{X}} c(y) P(x, dy), \quad (3)$$

$$\mu P(B) \triangleq \int_{\mathbb{X}} P(x, B) \mu(dx), \quad (4)$$

for $c \in B(\mathbb{X})$ and $\mu \in \mathcal{M}(\mathbb{X})$. We see that if $\mu$ is the distribution of $X_t$, then $\mu P$ is the distribution of $X_{t+1}$. Also,

$$Pc(x) = \mathbb{E}[c(X_{t+1})|X_t = x]. \quad (5)$$

The measure $\mu$ is an invariant measure of $P$ if

$$\mu = \mu P,$$

i.e: an invariant measure of $P$ is a fixed point of the (left) operator of $P$ in (4).

For a given weight function $u : \mathbb{X} \to [1, \infty)$ the $u$-norm of $P$ is defined as

$$||P||_u = \sup_{\mathbb{X}} u(x)^{-1} \int_{\mathbb{X}} u(y) |P(x, dy)|.$$

where $|P(x, .)|$ is the total variation of the measure $P(x, .)$. If for a $u$, $||P||_u < 1$, then the mappings defined in (3) and (4) are contractions.

The *Banach's Fixed Point Theorem* states that a contraction map on a complete metric space has a unique fixed point.

The *Mean Ergodic Theorem* states that, if $\mu$ is the unique invariant measure of $P$, then

$$c^* \triangleq \lim_{N \to \infty} \frac{1}{N} \sum_{t=0}^{N-1} P^t c = \int_{\mathbb{X}} c \, d\mu. \quad (6)$$

and $c^*$ is a constant ($\mu$ almost-everywhere).

For a given conditional probability $P$ on $\mathbb{X}$, and a fixed function $c : \mathbb{X} \to \mathbb{R}_+$, the system of equations

$$\begin{aligned} g + f &= c + Pf, \\ g &= Pg. \end{aligned} \quad (7)$$

is called the Poisson's equation. For a solution $(g, f)$, the function $g$ is called the invariant (or harmonic) function, and

$$g = \lim_{N \to \infty} \frac{1}{N} \sum_{t=0}^{N-1} P^t g \doteq \lim_{N \to \infty} \frac{1}{N} \sum_{t=0}^{N-1} P^t c,$$

where the second equality is true only if $P^n f / n \to 0$. If $\mu$ is the unique invariant measures of $P$, then $g = c^*$ in (6).

For the Markov decision process we consider this version of Poisson Equation ($\alpha$ as a constant)

$$\alpha + f(x) = \min_{a \in \mathcal{A}} [c(x) + P_a f(x)]. \quad (8)$$

The following theorem gives a condition for the optimal policy of a Markov decision process [6],[4].

*Theorem 1:* For the average cost MDP scheduling problem with cost function $c$, if $f$ solves Equation (8) for some constant $\alpha$, then the policy

$$w^*(x) = \arg\min_{a \in \mathcal{A}}[c(x) + P_a f(x)] \qquad (9)$$

is optimal conditioned that

$$\frac{1}{n} P_w^n f(x) \to 0, \quad n \to \infty. \qquad (10)$$

An algorithm for finding the optimal policy is the Policy Iteration Algorithm. This algorithm iteratively generates a sequence of policies $w_n$ starting from an initial policy $w_0$. Given $w_n$ in its $n$-th iteration, the algorithm finds $w_{n+1}$ by the following two steps.

- For the Markov chain with conditional probability $P_{w_n}$, solve equation (7) for $f_n$ (up to a constant).
- Find $w_{n+1}$ by $w_{n+1}(x) = \arg\min_{a \in \mathcal{A}}[c(x) + P_a f_n(x)]$.

It is shown that for finite state space $\mathbb{X}$, the sequence of policies $w_n$ converges to the optimal policy satisfying the optimality condition (9). For general state spaces sufficient conditions for convergence has been given in [6].

## III. THE OPTIMAL OBSERVABILITY PROBLEM

Consider a pair of correlated processes $(\mathbf{S}, \mathbf{Z})$ with finite domain sets $\mathcal{S}$ and $\mathcal{Z}$, respectively. We define two random vectors $\pi_n$ and $\rho_n$ as functions of $Z^{n-1}$ on the domains $\nabla_\mathcal{S}, \nabla_\mathcal{Z}$, respectively,

$$\pi_n(Z^{n-1}) = \underline{p}(S_n|Z^{n-1}). \qquad (11)$$

$$\rho_n(Z^{n-1}) = \underline{p}(Z_n|Z^{n-1}). \qquad (12)$$

According to our notation, the random vector $\pi_n$ (similarly $\rho_n$) has elements $\pi_n[k] = p(S_n = k|Z^{n-1})$, $k = 1, 2, ..., |\mathcal{S}|$.

For a hidden Markov process it is shown [7] that there are mappings $\zeta : \nabla_\mathcal{S} \to \nabla_\mathcal{Z}$ and $\eta : \mathcal{Z} \times \nabla_\mathcal{S} \to \nabla_\mathcal{S}$ such that for any $n$,

$$\begin{aligned} \rho_n &= \zeta(\pi_n) \\ \pi_{n+1} &= \eta(z_n, \pi_n), \end{aligned} \qquad (13)$$

The results on the analysis and representation of such a process can be extended to any pair of joint processes for which the mapping $\zeta$ and $\eta$ exist irrespective of the map definitions. The existence of such mappings in fact implies that such a joint process can be described by an iterated function system [8],[7]. Therefore we take the virtue of the existences of mappings (13) as the core property of special group of joint processes that we call *H-processes*. In this paper, this definition helps to bridge the partially observed Markov decision problems to simpler Markov decision problems.

*Definition 1:* A pair of correlated processes $(\mathbf{S}, \mathbf{Z})$ is called an H-process if the sequences $\pi_n$ and $\rho_n$ are related by some mappings $\zeta$ and $\eta$ as in (13).

We refer to $\mathbf{S}$ as the hidden component and $\mathbf{Z}$ as the observable component of the H-process. An example of H-process is the hidden Markov process.

A key property of H-process is that $\pi_n$ is a Markov chain on $\nabla_\mathcal{S}$ with the transition probability $P(x, B)$ defined by

$$P(x, B) = \sum_{l=1}^{|\mathcal{Z}|} 1_B(\eta(l, x))\zeta(x)[l], \qquad (14)$$

(where $1_B(.)$ is the set identity function), or equivalently by

$$P(x, x') = \begin{cases} (\zeta(x))[l] & , x' = \eta(l, x), \quad l = 1, 2, ..., |\mathcal{Z}| \\ 0 & \text{otherwise.} \end{cases} \qquad (15)$$

In fact the left operation of $P$ in (4) represents the evolution of 'distribution of $\pi_n$'(denoted by $\mu_n$) as probability measures on $\nabla_\mathcal{S}$, i.e: $\mu_{n+1} = \mu_n P$, which is easy to verify by the iterated function system representation of such processes [7, Eq. 1].

The controlled version of H-process can be described by considering an action process $A_n \in \mathcal{A}$ ($\mathcal{A}$ is the control set) and the existence of mappings $\zeta_a$ and $\eta_a$, $\forall a \in \mathcal{A}$, such that

$$\begin{aligned} \rho_n &= \zeta_{a_n}(\pi_n), \\ \pi_{n+1} &= \eta_{a_n}(z_n, \pi_n). \end{aligned} \qquad (16)$$

A controlled H-process defines a Markov decision process on $\nabla_\mathcal{S}$ with action set $\mathcal{A}$, and conditional probability for each action $a$ as

$$P_a(x, B) = \sum_{l=1}^{|\mathcal{Z}|} 1_B(\eta_a(l, x))\zeta_a(x)[l]. \qquad (17)$$

For a controlled H-process a stationary control policy is a function $w : \nabla_\mathcal{S} \to \mathcal{A}$ which deterministically connects $A_n$ to $\pi_n$, $A_n = w(\pi_n)$. The pair $(\mathbf{S}, \mathbf{Z})$ when controlled by a policy $w$ is an H-process with mappings,

$$\begin{aligned} \zeta_w(x) &= \zeta_{w(x)}(x) \\ \eta_w(z, x) &= \eta_{w(x)}(z, x). \end{aligned} \qquad (18)$$

Hence, a controlled H-process with policy $w$ defines a Markov chain on $\nabla_\mathcal{S}$ with conditional probability

$$P_w(x, B) = P_{w(x)}(x, B).$$

Now using H-process, we show that the optimal observability problem of POMDP,s can be considered as finding an optimal policy for a Markov decision process.

A POMDP is a controlled version of a hidden Markov process (HMP). Consider an HMP [7] with state transition probability matrix $Q$, representing the dynamics of the process, and measurement (emission) matrix $T$, representing the sensor. A POMDP is an HMP when these matrices are functions of a control action $a \in \mathbb{A}$, and we chose the action based on our past and current observations for either controllability or observability purposes. In the controllability problem the aim is to control the state process to move it towards more favorite states, but in the observability problem the state process is autonomous and the aim is to dynamically adjust the measuring apparatus to have the best observation of the state process. In both problems the control action is based on $\pi_n$ which represents the belief on the state, and it is a sufficient statistics for past observations. A control policy is a rule for choosing actions based on this belief for prospective minimization of the average (or discounted) expectation of a cost function over belief space. For the control problem the cost function is $c(\pi) = \pi\beta'$, where $\beta \in \mathbb{R}^{|\mathcal{S}|}$ is a fixed vector (states' costs). Hence for the control problem the cost function is linear. This case is not true for the observability problem.

To formulate the observability problem as a Markov decision problem, first we consider that HMP is an H-process with the following mappings [7],

$$\zeta(\pi) = \pi T,$$

$$\eta(z, \pi) \triangleq \frac{\pi D(z) Q}{\pi D(z) \underline{1}}, \qquad (19)$$

where $D(z)$ is a diagonal matrix with $d_{k,k}(z) = T[k, z]$, $k = 1, 2, .., |\mathcal{S}|$. We also note that in the controllability problem it is (usually only) $Q$ that is a function of the action $a$, whereas in observability problem $T$ is a function of $a$. As a result in the controllability problem usually $\zeta$ is fixed but we have a set of functions $\eta_a$, similar to (19) with $Q(a)$ instead of $Q$. These define a controlled H-process which corresponds to an MDP. The linearity of cost function helps to solve this problem by value iteration algorithm using incremental pruning [5].

In contrast, for the observability problem the POMDP is a controlled H-process with the following mappings

$$\zeta_a(\pi) = \pi T(a),$$

$$\eta_a(z, \pi) \triangleq \frac{\pi D(z, a) Q}{\pi D(z, a) \underline{1}}. \qquad (20)$$

where $D(z, a)$ is a diagonal matrix with $d_{k,k}(z) = T(a)[k, z]$, $k = 1, 2, .., |\mathcal{S}|$. Moreover, the cost function on $\nabla_\mathcal{S}$ cannot be a linear function. The (positive) cost function needs to be designed in such a way that imposes higher cost when the belief $\pi_n$ moves away from the vertices of $\nabla_\mathcal{S}$. At the vertices, the belief about state is complete certainty, thus zero cost. A policy is optimal which ensures that as the state process evolves autonomously, the belief $\pi_n$ in its expectation hops only between the regions close to the vertices, so the average ambiguity about state (cost) is minimized.

In previous works [9],[2], the cost function $c(\pi) = 1 - \pi\pi'$ has been considered for this problem due to its ease of approximation by piecewise linear (and zero cost at vertices), hence allowing application of dynamic programming under discounted criterion. In this paper we consider the entropy function as the cost function, $c(\pi) = h(\pi)$, and average cost criterion, hence allowing application of PIA. An interesting result is that under ergodicity condition the average cost $J(w, x)$ is the same as the estimation entropy defined in [7] for the H-process $(\mathbf{S}, \mathbf{Z})$ corresponding to $w$. We first discuss this relation in the next section and then formalize the observability problem of sensor scheduling as an MDP scheduling problem.

## IV. THE ESTIMATION ENTROPY

The entropy rate of a process $\mathbf{Z}$ is defined as

$$\hat{H}_Z \triangleq \lim_{n \to \infty} \frac{1}{n} H(Z_0^n), \qquad (21)$$

when the limit exists. From chain rule, we can write

$$H(Z_i | Z_0^{i-1}) = H(Z_0^i) - H(Z_0^{i-1}).$$

We see that the entropy rate is the limit of Cesaro mean of the above $i$-sequence, i.e:

$$\hat{H}_Z = \lim_{n \to \infty} \frac{1}{n} \sum_{i=1}^{n} H(Z_i | Z_0^{i-1}). \qquad (22)$$

For an H-process we define the entropy rate as the entropy rate of its observable component. Along the line of (22), the estimation entropy [7] for an H-process is defined as

$$\hat{H}_{S/Z} \triangleq \lim_{n \to \infty} \frac{1}{n} \sum_{i=1}^{n} H(S_i | Z_0^{i-1}). \qquad (23)$$

We know that if a sequence $\alpha_n$ converges, then the sequence of its Cesaro mean (i.e: $1/n \sum^n \alpha_i$) also converges to the same limit [10, Theorem 4.2.3]. As a result for stationary H-processes, the entropy rate and estimation entropy can be written as

$$\begin{aligned} \hat{H}_Z &= \lim_{n \to \infty} H(Z_n | Z_0^{n-1}), \\ \hat{H}_{S/Z} &= \lim_{n \to \infty} H(S_n | Z_0^{n-1}). \end{aligned} \qquad (24)$$

From its definition we see that the estimation entropy for an H-process is the limit of running average of residual uncertainty about the hidden component under the knowledge of all past observed process, thus it inversely measures the observability of the hidden process. However we also show that under ergodicity conditions the estimation entropy is the long run average entropy of the belief process. To this end, using the relations in section II, we can prove the following interesting results [11].

*Lemma 1:* For an H-process,

$$H(S_n | Z_0^{n-1}, \pi_0 = x) = P^n h(x), \qquad (25)$$

where $P$ is defined by (14) and $h$ is the entropy function.

*Theorem 2:* For an H-process, if $P$ has unique invariant measure $\mu$, then

$$\hat{H}_{S/Z} = \lim_{n \to \infty} \frac{1}{n} \sum_{t=0}^{n-1} P^t h = \int_{\nabla_\mathcal{S}} h \, d\mu. \qquad (26)$$

For a hidden Markov process which has primitive matrix $Q$ and also matrix $T$ has nonzero elements, it is shown that the integral expression in (26) for $\hat{H}_{S/Z}$ is true for any attractive and invariant measure $\mu$ [7, Theorem 1].

According to (5) and Theorem 2, we see that under ergodicity condition (existence of a unique invariant measure) the estimation entropy of an H-process is the average cost of a

Markov chain with conditional probability $P$ in (14) and the cost function as the entropy function $h$,

$$\hat{H}_{S/Z} = \lim_{n \to \infty} \frac{1}{n} \sum_{t=0}^{n-1} \mathbb{E}[h(X_t)|X_0 = x] = J(x). \quad (27)$$

For a controlled H-process the estimation entropy as a function of policy $w$ is this average cost for the Markov chain corresponding to $P_w$, i.e: $\hat{H}_{S/Z}(w) = J(w,x), \forall x$, provided that $P_w$ has a unique invariant measure.

Using the Banach's fixed point Theorem, we see that if $||P_w||_u < 1$ for some $u : \nabla_{\mathcal{S}} \to [1, \infty)$, then $P_w$ has unique invariant measure. By this we infer that a sufficient condition for the existence of a unique invariant measure of $P_w$ is that for some $u : \nabla_{\mathcal{S}} \to [1, \infty)$,

$$\sum_z u(\eta_w(z,x))\zeta_w(x)[z] < u(x), \qquad \forall x \in \nabla_{\mathcal{S}}. \quad (28)$$

Although according to (27) under ergodicity condition (for any $w$) the POMDP observability problem for minimum estimation entropy $\hat{H}_{S/Z}(w)$ is equivalent to the average cost MDP scheduling problem for $J(w,x)$, we consider the sensor scheduling problem only as the average cost MDP problem. This MDP is defined by the set of conditional probabilities $P_a$ in (17), where $\zeta_a$ and $\eta_a$ are defined by (20). In a similar problem, minimum estimation entropy criterion has been considered for finding the optimal policy of multiple measurement hidden Markov processes [12]. This analysis required the ergodicity of $P_w$ for any $w$. Under such a condition, for that problem an optimality criterion simpler than Theorem 1 has been conjectured.

## V. Sensor Scheduling for Optimal Observability

Using the POMDP observability problem as the framework for sensor scheduling, here we formalize this scheduling as an average cost MDP scheduling problem. Let $M, L, A$ be the cardinality of state, measurement, and control sets, respectively. The integer $A$ represents the number of sensors. We also have a $M \times M$ state transition probability matrix Q, and a set of $M \times L$ measurement (sensor) probability matrices $T_a$, a=1,2,...,A. The description of MDP scheduling framework based on [6] for the sensor scheduling problem is as follows:

- The Markov decision process evolves on the state space $\mathbb{X}$, where $\mathbb{X}$ is the probability simplex in $\mathbb{R}^M$.
- The action set is $\mathcal{A} = \{1, 2, ...A\}$, and the admissible action set for any $x \in \mathbb{X}$ is $\mathcal{A}$.
- The set of conditional probability distributions $P_a(x, B), a \in \mathcal{A}$, is defined by

$$P_a(x, B) = \sum_{l=1}^{|\mathcal{Z}|} 1_B(\eta_a(l, x))(xT_a)[l]. \quad (29)$$

where

$$\eta_a(l, x) \triangleq \frac{xD_a(l)Q}{xD_a(l)\underline{1}}, \quad (30)$$

and $D_a(l)$ is a diagonal matrix with $m$-th diagonal element being $T_a[m,l]$, $m = 1, 2, .., M$.

- For any stationary policy $w : \mathbb{X} \to \mathcal{A}$ the state process $\mathbf{X}(w) = \{X_t(w) : t \in \mathbb{Z}\}$ is a Markov chain with conditional probability $P_w(x, B) = P_{w(x)}(x, B)$.
- The average cost of a policy $w$ for a given initial condition $x$ is

$$J(w, x) = \limsup_{N \to \infty} \frac{1}{N} \sum_{t=0}^{N-1} \mathbb{E}_x[c(X_t(w))], \quad (31)$$

where the cost function $c : \mathbb{X} \to [0, \log(M)]$ is the entropy function defined by

$$c(x) = h(x) \triangleq -\sum_m x[m] \log x[m].$$

Objective: Find the optimal policy $w^*$ where $J(w^*, x) \leq J(w, x)$ for all polices $w$ and any initial state $x$.

As an average cost MDP scheduling problem the objective can be achieved by policy iteration algorithm [6]. The solution $w^*$ will then be the optimum observation policy. Future projection is an adaptation of PIA and a rigorous analysis of convergence for the algorithm under the conditions of this problem, using the results of [6].

## VI. Acknowledgment

This work was supported in part by the Defense Advanced Research Projects Agency of the US Department of Defense and was monitored by the Office of Naval Research under Contract No. N00014-04-C-0437.

The conjecture on a simpler optimality criterion quoted from the joint work of [12] is due to Bill Moran and Sofia Suvorova.

## References

[1] D. J. Kershaw and R. J. Evans. Waveform selective probabilistic data association. *IEEE Trans. on Aerospace and Electronic Systems*, 33(4):1180–1188, 1997.

[2] B. La Scala, M. Rezaeian, and B. Moran, "Optimal Adaptive Waveform Scheduling for Target Tracking", in proceedings of International Conference on Information Fusion, July 2005, Philadelphia, PA, USA.

[3] R.D. Smallwood and E.J. Sondik, "Optimal control of partially observed Markov processes over a finite horizon," *Operation Research*, vol.21, pp. 1071-1088, 1973.

[4] O. Hernandez, J. Lasserre, *Discrete Time Markov Control Processes*, volumes 1 and 2, Springer-Verlag, 1996 and 1999.

[5] W.S. Lovejoy, "A survey of algorithmic methods for partially observed Markov decision processes", *Operation Research*, vol.39, no. 1, pp. 162-175, Jan 1991.

[6] S. Meyn, " The policy iteration algorithm for average reward Markov decision processes with general state space," *IEEE Trans. Automatic Control*, vol.42 No.12 , pp. 1663–1679, Dec. 1997.

[7] M. Rezaeian, "Hidden Markov process: a new representation, entropy rate and estimation entropy," submitted to *IEEE Trans. Inform. Theory*, June 2006, available at http://arxiv.org/cs.IT/0606114 .

[8] W. Slomczynski. "Dynamical Entropy, Markov Operators and Iterated Function Systems", Wydawnictwo Uniwersytetu Jagiellonskiego, ISBN 83-233-1769-0, Krakow, 2003.

[9] V. Krishnamurthy, " Algorithms for optimal scheduling and management of hidden Markov model sensors ," *IEEE Trans. Signal Processing*, vol.50 No.6 , pp. 1382–1396, June 2002.

[10] T.M.Cover and J.A.Thomas. "Elements of Information Theory", Wiley, New York, 1991.

[11] M. Rezaeian, "The Optimal Observability of Partially Observable Markov Decision Processes ," In preparation for journal publication.

[12] –, "Minimum entropy scheduling for hidden Markov processes," *Raytheon Systems Company* internal report, Integrated Sensing Processor Phase II. March 2006.